\documentclass[12pt]{iopart}

\usepackage{iopams}  
\usepackage{graphicx,xcolor,float}
\begin{document}

\title[Metamaterials designed for enhanced ENZ properties]{Metamaterials designed for enhanced ENZ properties}

\author{Matias Koivurova$^1$, Tommi Hakala$^2$, Jari Turunen$^2$, Ari T. Friberg$^2$, Marco Ornigotti$^1$, Humeyra Caglayan$^1$}

\address{$^1$Faculty of Engineering and Natural Sciences, Photonics, Tampere University, 33720 Tampere, Finland}
\address{$^2$Institute of Photonics, University of Eastern Finland, P.O. Box 111, FI-80101 Joensuu, Finland}

\ead{matias.koivurova@tuni.fi}
\vspace{10pt}

\begin{abstract}
We examine layered metamaterial structures consisting of alternating films of epsilon-near-zero (ENZ) and dielectric material, and show that for such a stack it is possible to enhance the refractive, reflective or absorptive properties of the ENZ. The proposed structure takes advantage of resonances from several interfaces, guided modes, and plasmon excitations to achieve the desired enhancement, and it is not an effective medium. We use analytical modeling tools to show how the different degrees of freedom affect the properties of the stack, and propose experimentally feasible parameters for such structures.
\end{abstract}

%
\vspace{2pc}
%
\submitto{\NJP}
%
%
%

\section{Introduction}

In recent years, materials with unconventional macroscopic properties have become more and more important in the study of optics and photonics. One such class of materials are epsilon-near-zero (ENZ) media. The unique and fascinating features of ENZ structures enable the realization of advanced optical applications, such as directional light enhancement \cite{7Hajian,8Hajian}, coherent perfect absorption \cite{9Feng}, radiation pattern tailoring \cite{10Alu2007}, nonlinear fast optical switching \cite{11Wurtz}, and high speed tunable devices \cite{Alam, Alireza}. Although the aforementioned properties and features of ENZ have been demonstrated for the case of bulk materials, we are still relatively far from having all of the possible breakthroughs in practical experimental settings. Additionally, some theoretical criticism regarding the usefulness of ENZ materials in certain tasks has surfaced recently \cite{realandimag}.

The term ENZ is often understood to imply either that only the real part of the dielectric constant is close to zero, or that both real and imaginary parts are almost zero. However, the latter definition automatically means that the material is a zero index medium \cite{zim} and -- maybe less obviously -- that no electromagnetic energy can be coupled into it \cite{realandimag}. On the other hand, the former definition fundamentally implies that ENZ materials have a finite imaginary part, and thus better describes ENZs that can be fabricated and characterised experimentally. The relatively large imaginary part constitutes one of the main problems in current ENZ research, and for this reason we will focus our attention on this definition. Although this class of materials has been predicted to exhibit interesting refractive properties, employing them is experimentally challenging, due to the associated high losses. To mitigate this effect, one may be tempted to use only a sub-wavelength film of the material, but then the refractive properties would essentially be lost. In fact, if a plane wave impinges upon a thin film of ENZ at an angle larger than the critical angle, the evanescent field can tunnel through the film and a large portion of the energy will be transmitted. Thus, currently available ENZ materials cannot be used to realize their full potential.

In the present study, we examine layered metamaterial structures consisting of alternating thin films of ENZ and dielectric. We show that with these designs, it is possible to attain highly directional transmission (or reflection), far beyond the capability of the ENZ material alone. In addition to strongly enhanced directionality, the stack can suppress overall losses to only a fraction when compared to the employed ENZ material alone. Moreover, the metamaterial can also be designed to feature very high absorption, on par with the most strongly absorbing novel materials, although our stack design is perfectly planar (no surface modification). The proposed structures take advantage of Fabry-P\'erot resonances, guided modes, and plasmon excitations to achieve the observed enhancements. Most importantly, the designed metamaterials are \textit{not} effective ENZ media, since the film thicknesses are either too large for effective approximation, or the effective permittivities are nowhere near the ENZ regime. Therefore, it is more appropriate to call them enhanced ENZs, rather than effective ENZs. We employ the theory of stratified media \cite{stratified} to analytically model how the different degrees of freedom affect the refraction and reflection properties of the stack, and propose experimentally feasible parameters for such structures.

The present work is organised as follows: in section 2, we go through the theory of stratified media, which was used to calculate the Fresnel coefficients employed in section 3. In subsections 3.1, 3.2, and 3.3, we discuss optimizations for transmission, reflection, and absorption, respectively. Section 4 contains concluding remarks and a summary of our main findings.

\section{Modeling method}

\begin{figure}[H]
\centering
\includegraphics[width=0.5\columnwidth]{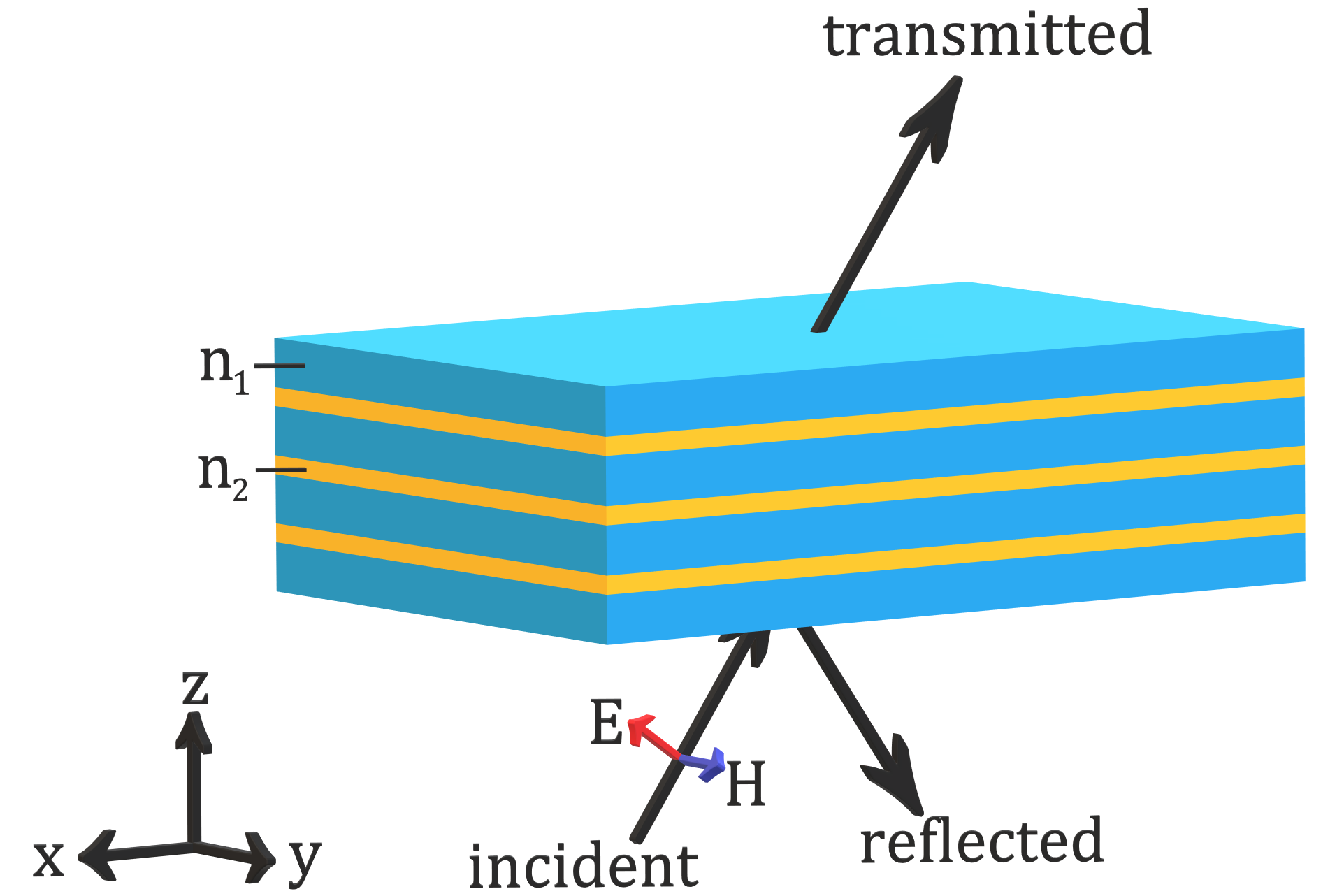}
\caption{Schematic representation of the proposed structure with two different materials and TM polarized incident light.}
\label{structure}
\end{figure}

Let us consider a structure of parallel planar layers of homogeneous media as in Fig.~\ref{structure}, where we have two different materials (though their number is not limited). The refractive indices of these layers -- $n_1$ and $n_2$ -- can be either real or complex, allowing for the modeling of arbitrary material stacks. We will consider transverse magnetic (TM) and transverse electric (TE) polarizations separately.

In the following, we will consider a two dimensional geometry along the $x$-$z$-plane and assume invariance along the $y$-axis, with the axes fixed as in Fig.~\ref{structure}. The electric field at the front surface of a single layer can be connected to the field at its back surface via a simple $2 \times 2$ transfer matrix of the form

\begin{equation}
\textbf{M}^{(l)}_{\tiny{\textrm{TM}}} = \left[
    \begin{array}{cc}
            \cos(\beta_l d_l)       & \sin(\beta_l d_l)n_l^2/\beta_l \\
    -\sin(\beta_l d_l)\beta_l/n_l^2 &       \cos(\beta_l d_l)        \\
    \end{array}\right],
    \end{equation}
for TM polarized light and 
\begin{equation}
    \textbf{M}^{(l)}_{\tiny{\textrm{TE}}} = \left[
    \begin{array}{cc}
            \cos(\beta_l d_l)       & \sin(\beta_l d_l)/\beta_l \\
    -\sin(\beta_l d_l)\beta_l &       \cos(\beta_l d_l)        \\
    \end{array}\right],
\end{equation}
For TE polarization. Here $\beta_l = \sqrt{k_0^2 n_l^2 - k_x^2}$ is the propagation constant inside the slab, $k_0$ being the vacuum wavenumber, $k_x = k_0\sin\theta_i$ is the transverse wavevector component, and $\theta_i$ is the angle of incidence. Moreover, $n_l$ is the refractive index of the material, and $d_l$ the thickness of the layer.

Since both the electric field and its $z$-derivative are continuous at each interface, we can form a chain of $L$ layers stacked on top of each other, by simply multiplying together the transfer matrices of each single layer as
\begin{equation}
    \textbf{M} = \textbf{M}^{(L)}\textbf{M}^{(L-1)}...\textbf{M}^{(2)}\textbf{M}^{(1)},
\end{equation}
for either polarization. Taking the inverse matrix $\textbf{M}^{-1}(i,j)$, where $i$ and $j$ denote the appropriate matrix elements, we may write the extended Fresnel equations, i.e.,

\begin{equation}
    \hspace{-20mm} r_{\tiny{\textrm{TM}}} = - \frac{\textbf{M}^{-1}_{\tiny{\textrm{TM}}}(1,1) \beta_i n_t^2 - \textbf{M}^{-1}_{\tiny{\textrm{TM}}}(2,2) \beta_t n_i^2 + i(\textbf{M}^{-1}_{\tiny{\textrm{TM}}}(1,2) n_i^2n_t^2 + \textbf{M}^{-1}_{\tiny{\textrm{TM}}}(2,1)\beta_i\beta_t)}{\textbf{M}^{-1}_{\tiny{\textrm{TM}}}(1,1) \beta_i n_t^2 + \textbf{M}^{-1}_{\tiny{\textrm{TM}}}(2,2) \beta_t n_i^2 - i(\textbf{M}^{-1}_{\tiny{\textrm{TM}}}(2,1) n_i^2n_t^2 - \textbf{M}^{-1}_{\tiny{\textrm{TM}}}(2,1)\beta_i\beta_t)},
\end{equation}
and
\begin{equation}
    \hspace{-20mm} t_{\tiny{\textrm{TM}}} = - \frac{2n_in_t\beta_i}{\textbf{M}^{-1}_{\tiny{\textrm{TM}}}(1,1) \beta_i n_t^2 + \textbf{M}^{-1}_{\tiny{\textrm{TM}}}(2,2) \beta_t n_i^2 - i(\textbf{M}^{-1}_{\tiny{\textrm{TM}}}(2,1) n_i^2n_t^2 - \textbf{M}^{-1}_{\tiny{\textrm{TM}}}(1,2)\beta_i\beta_t)}.
\end{equation}
The corresponding coefficients for transverse electric (TE) polarized light have a similar form,
\begin{equation}
    r_{\tiny{\textrm{TE}}} = - \frac{\textbf{M}^{-1}_{\tiny{\textrm{TE}}}(1,1) \beta_i - \textbf{M}^{-1}_{\tiny{\textrm{TE}}}(2,2) \beta_t + i(\textbf{M}^{-1}_{\tiny{\textrm{TE}}}(2,1) + \textbf{M}^{-1}_{\tiny{\textrm{TE}}}(1,2) \beta_i\beta_t)}{\textbf{M}^{-1}_{\tiny{\textrm{TE}}}(1,1) \beta_i + \textbf{M}^{-1}_{\tiny{\textrm{TE}}}(2,2) \beta_t - i(\textbf{M}^{-1}_{\tiny{\textrm{TE}}}(2,1) - \textbf{M}^{-1}_{\tiny{\textrm{TE}}}(1,2) \beta_i\beta_t)},
\end{equation}
and
\begin{equation}
    t_{\tiny{\textrm{TE}}} = - \frac{2\beta_i}{\textbf{M}^{-1}_{\tiny{\textrm{TE}}}(1,1) \beta_i + \textbf{M}^{-1}_{\tiny{\textrm{TE}}}(2,2) \beta_t - i(\textbf{M}^{-1}_{\tiny{\textrm{TE}}}(2,1) - \textbf{M}^{-1}_{\tiny{\textrm{TE}}}(1,2) \beta_i\beta_t)}.
\end{equation}
The subscripts $i$ and $t$ refer to incident and transmitted sides of the stack, respectively. We consider the stack to be surrounded by air, and thus $n_i = n_t \approx 1$. The intensity reflection and transmission coefficients for both polarizations are obtained with $R = |r|^2$ and $T = \mathcal{R}\left\{\beta_t/\beta_i\right\}|t|^2$, where $\mathcal{R}$ denotes the real part (here $\beta_t/\beta_i = 1$). Absorption $A$ can be found by assuming energy conservation, such that $T + R + A = 1$.

The materials we choose to model are indium tin oxide (ITO), which is highly dispersive around the ENZ wavelength, and titanium dioxide (TiO$_2$) that has very weak dispersion. The dispersion of ITO was accounted for by using the Drude model,
\begin{equation}
\epsilon = \epsilon_{\infty} - \frac{\omega_p^2}{\omega^2 + i\,\Gamma\,\omega},
\end{equation}
with the parameters $\omega_p = 2.65 \times 10^{15}$ rad/s, $\Gamma = 2.05 \times 10^{14}$ rad/s, and $\epsilon_\infty = 3.91$, for plasma frequency, damping, and permittivity in the limit of infinite frequency, respectively \cite{Humeyra}. These choices are valid at least between a wavelength range of 1000 to 3000 nm, and lead to an ENZ wavelength of $\lambda = 1422$ nm, at which the permittivity of ITO is purely imaginary, $\epsilon_{1} \approx i\,0.61$. The real part of the resulting index of refraction ($n_1\approx0.55$) leads to a few noteworthy properties. 

Light at the ENZ wavelength that is incident from air onto ITO will experience a phenomenon similar to total internal reflection (TIR), since the light will be moving from higher index (air) to a lower index (ITO) material. Since the reflection occurs at the outer surface and the evanescent tail will experience loss, it may be more appropriate to describe it as maximal external reflection (MER) instead, although the angle can be found similarly to the familiar TIR critical angle. The critical angle at an air--ITO interface is 33.4$^\circ$, and above this it is possible to excite plasmons. Additionally, -- if we ignore losses for a moment -- the Brewster angle for this interface is quite close to the critical angle, at 28.7 degrees. Of course, due to the losses the standard definition is not valid and one should be considering pseudo-Brewster's angle instead, which is defined as the angle where TM polarized light experiences minimum reflection. It generally occurs at a slightly different angle of incidence than Brewster's angle, and finding it theoretically is far more challenging.

We set the incident wavelength to 1422 nm, and the permittivity of TiO$_2$ to a constant $\epsilon_{2} = 6.25$. Furthermore, we chose to have a relatively thick stack with 16 ENZ layers and 15 dielectric layers, starting and ending in ITO, with TiO$_2$ in between. It is to be expected that the desired properties become more apparent as the number of layers increases, although the overall losses will also increase. Thus, the number of layers is a compromise between the desired bulk properties and the optical density of the structure. What is left to be determined, are the optimal film thicknesses.

\section{Optimization}

The optimization is performed with a parameter sweep, first by fixing the ITO film thickness, $d_1$, to some arbitrary value (for example 100 nm), and letting the TiO$_2$ thickness, $d_2$, attain a wide range of values, so that we could obtain $(\theta_i,d_2)$-dependent maps for $T$, $R$, and $A$. We build the stack from layers of constant thickness, such that every ITO and TiO$_2$ layer has a thickness of $d_1$ and $d_2$, respectively. After finding the value that maximizes the attribute we are optimizing ($T$, $R$, or $A$), we keep that value and in turn produce $(\theta_i,d_1)$-dependent maps to update the ENZ thickness. After a few rounds of this process, we arrive at good estimates for the optimum film thicknesses. Once the film thicknesses are fixed, we scan the wavelength of the incident light to find the spectral properties of the stack within the wavelength range of 1000 to 2000 nm.

\subsection{Transmission}

\begin{figure}[H]
\centering
\includegraphics[width=0.9\columnwidth]{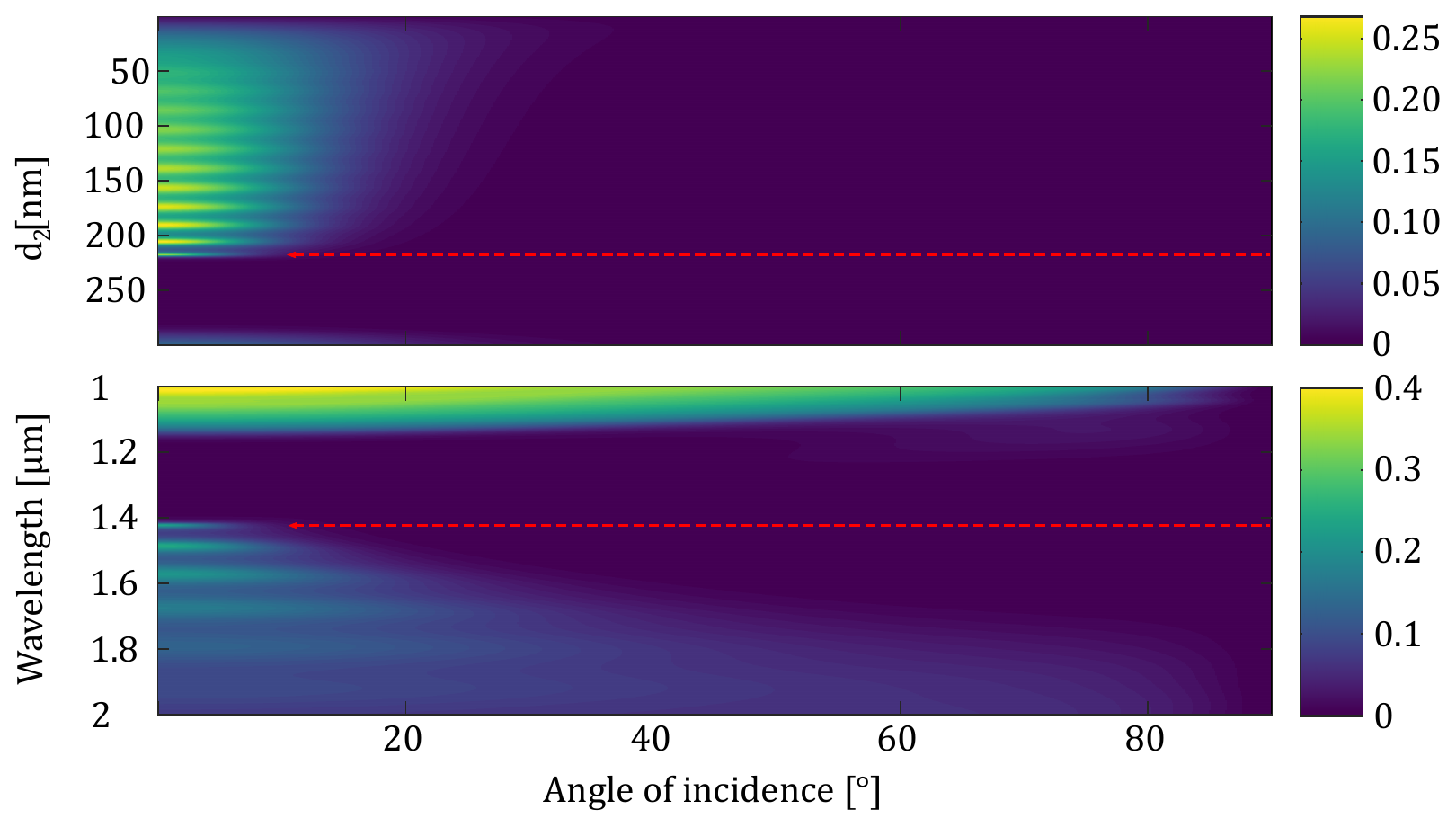}
\caption{Incident light is TM polarized, top: transmission as a function of angle of incidence, $\theta_i$, and dielectric film thickness, $d_2$. Bottom: transmission as a function of angle of incidence and wavelength, at the optimal dielectric thickness, $d_2=218$ nm. Both have 16 layers of ITO with a thickness of $d_1=64$ nm, and 15 layers of TiO$_2$.}
\label{transmission}
\end{figure}

The final $(\theta_i,d_2)$ dependent transmission map for TM polarized light is shown in Fig.~\ref{transmission} together with the wavelength dependence. The ITO thickness, $d_1$, is optimized to 64 nm and the optimal TiO$_2$ thickness, as well as the operating wavelength, are marked with red dashed lines.

As can be seen from the figure, there are several values of dielectric thickness where a resonance is attained, coinciding with the multiple reflections from different interfaces with a total of $N-1$ resonance peaks, where $N$ is the number of ENZ layers. They are grouped together inside larger resonance modes, which have a periodicity of about 285 nm, corresponding to $\lambda/2\sqrt{\epsilon_2}$. Curiously, the sharpest transmission peak is not obtained at any integer fraction of the wavelength. Instead, the optimum thickness for the TiO$_2$ is 218 nm. At this dielectric thickness, the spread of the transmitted light has a half-width at half-maximum (HWHM) of 4.6$^\circ$, with 23.2 $\%$ of normally incident light being transmitted. The angular width is on par with the current state-of-the-art directional matematerial devices, which feature angular spreads of about 10 degrees in full width \cite{zim,directional,Moitra2013,review}. The wavelength dependence shows multiple transmission peaks as well, implying that the structure is wavelength tunable.

\begin{figure}[H]
\centering
\includegraphics[width=0.9\columnwidth]{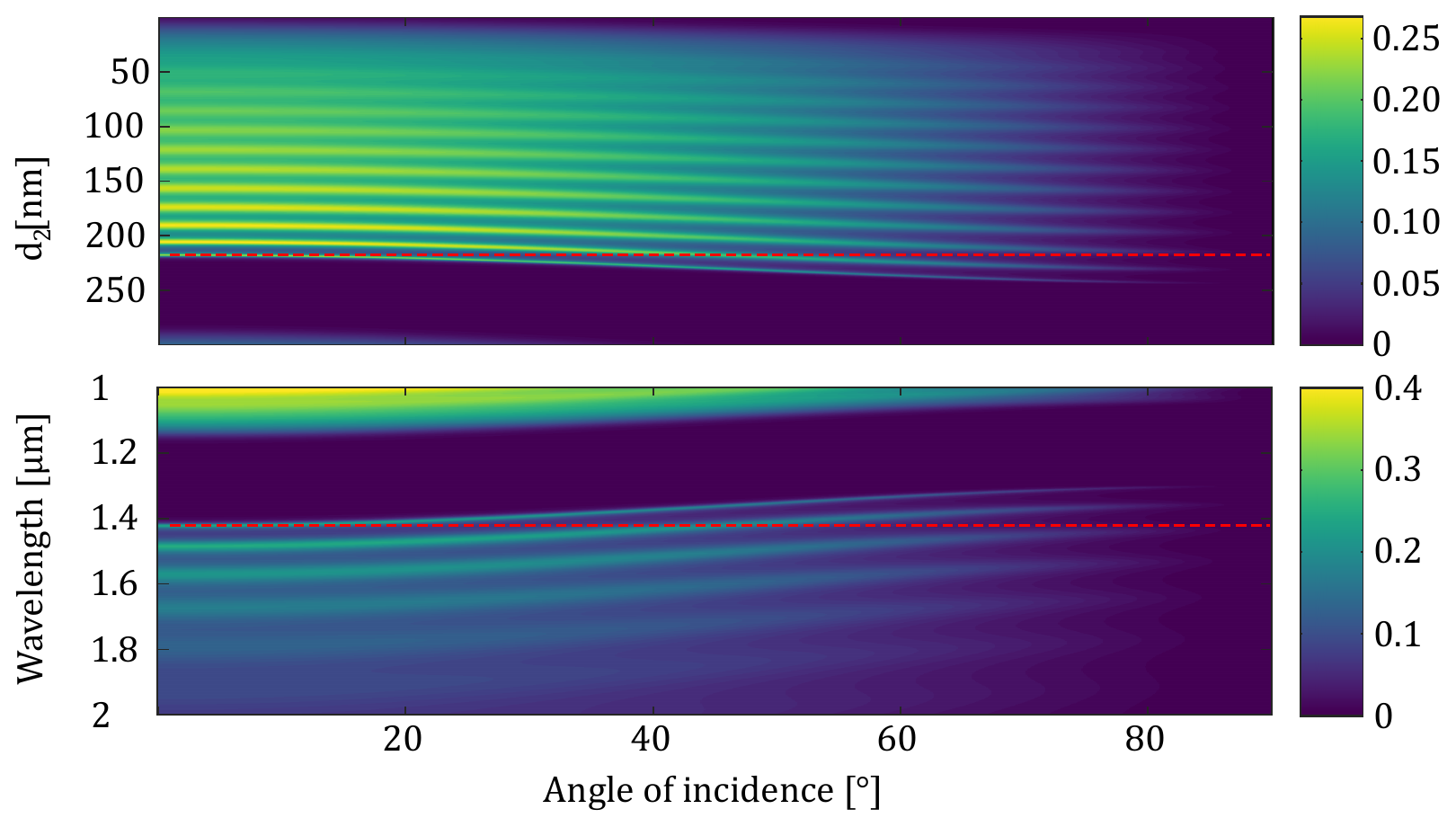}
\caption{Incident light is TE polarized, top: transmission as a function of angle of incidence, $\theta_i$, and dielectric film thickness, $d_2$. Bottom: transmission as a function of angle of incidence and wavelength, at the optimal dielectric thickness, $d_2=218$ nm. Both have 16 layers of ITO with a thickness of $d_1=64$ nm, and 15 layers of TiO$_2$.}
\label{transmission_TE}
\end{figure}

Similar maps for TE polarization are found in Fig.~\ref{transmission_TE}. Looking at the optimized TiO$_2$ thickness of 218 nm, the transmission cone is much wider, with a 11.5 degree HWHM, and there is an additional transmission peak at 45.2$^\circ$. The optimized structure transmits a maximum of $\sim$ 23 \% of incident light for both polarizations, but there is a large amount of absorption present in the TM case, which removes the unwanted side maxima.

The stack behaves as an angular pinhole, and obviously the performance is greatest with TM polarized input light. Recreating a 4.6$^\circ$ HWHM transmission cone with a low index material would require a refractive index of $n = 0.08$, which is far below the ITO index. Additionally, a single slab of ITO with the same thickness as all 16 ENZ layers put together -- i.e. a total thickness of $1064$ nm -- would transmit only $\sim$0.9 $\%$ of normally incident light. Thus, the designed metamaterial constitutes nearly a 26 fold increase in transmission, when compared to a bare ITO slab.

\subsection{Reflection}

\begin{figure}[H]
\centering
\includegraphics[width=0.9\columnwidth]{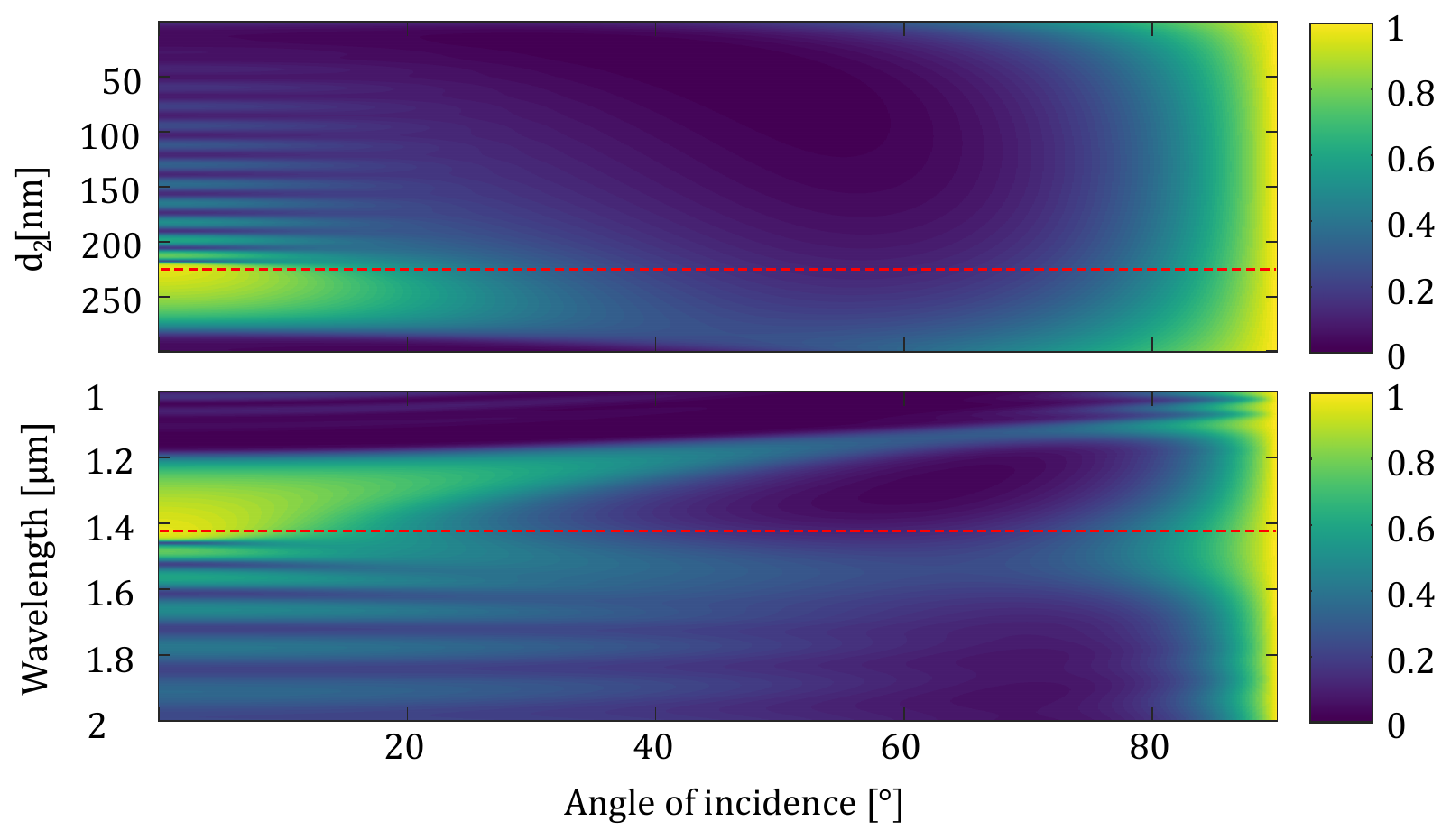}
\caption{Incident light is TM polarized, top: reflection as a function of angle of incidence, $\theta_i$, and dielectric film thickness, $d_2$. Bottom: reflection as a function of angle of incidence and wavelength, at the optimal dielectric thickness, $d_2=225$ nm. Both have 16 layers of ITO with a thickness of $d_1=64$ nm, and 15 layers of TiO$_2$}
\label{reflection}
\end{figure}

\begin{figure}[H]
\centering
\includegraphics[width=0.9\columnwidth]{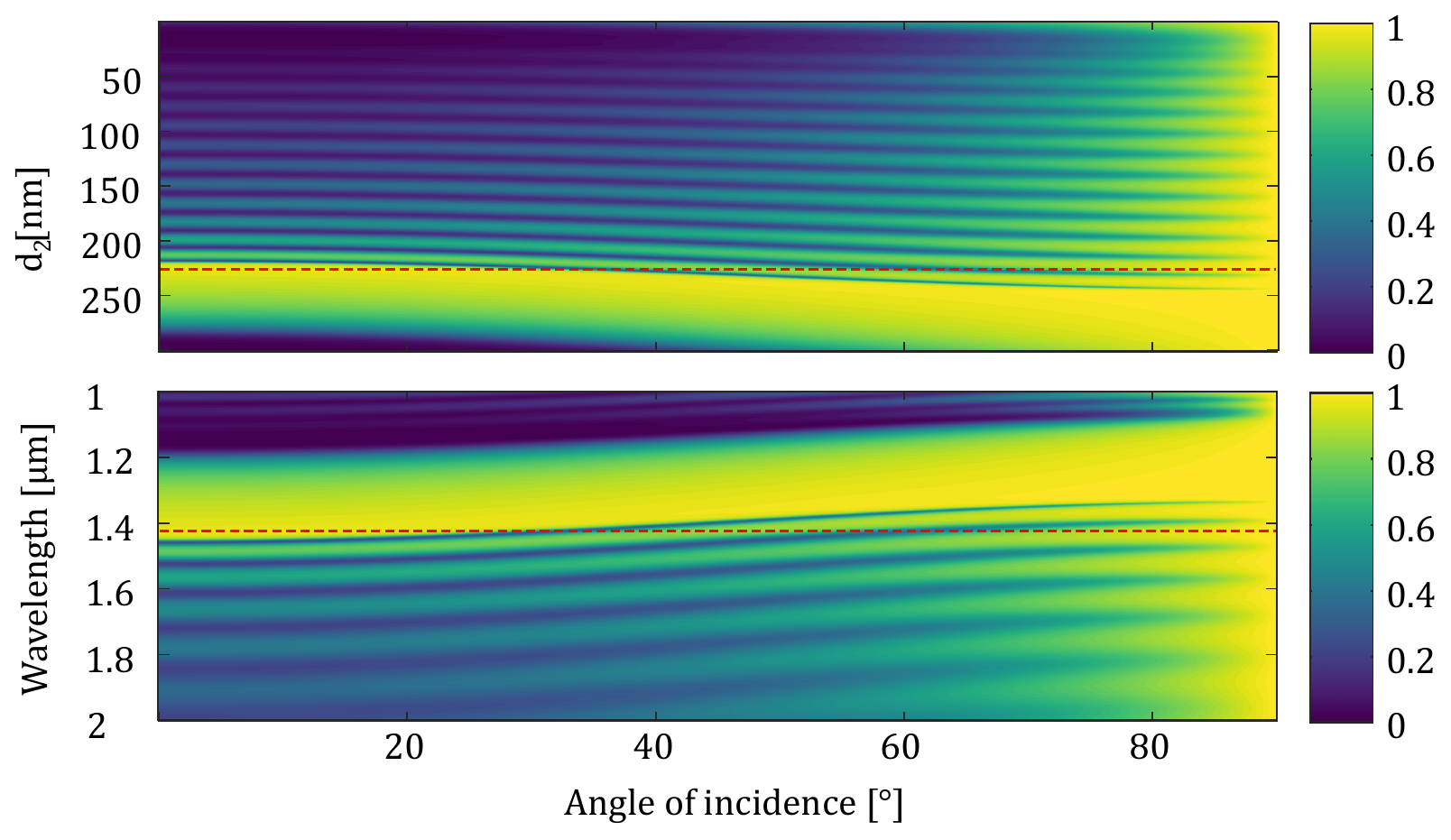}
\caption{Incident light is TE polarized, top: reflection as a function of angle of incidence, $\theta_i$, and dielectric film thickness, $d_2$. Bottom: reflection as a function of angle of incidence and wavelength, at the optimal dielectric thickness, $d_2=225$ nm. Both have 16 layers of ITO with a thickness of $d_1=64$ nm, and 15 layers of TiO$_2$}
\label{reflection_TE}
\end{figure}

Similarly to the transmission optimized case, the final $(\theta_i,d_2)$ dependent reflection map for TM polarization is shown in Fig.~\ref{reflection}, where the ITO thickness is still 64 nm and the optimal TiO$_2$ thickness is marked with a red dashed line. The wavelength dependence of the optimized structure is again found below the $(\theta_i,d_2)$ map, and the operating wavelength is also marked. The corresponding data for TE polarization can be found in Fig.~\ref{reflection_TE}.
 
The reflection was optimized with almost the same parameters as transmission, with ITO thickness remaining the same and the TiO$_2$ being only 7 nm thicker, at 225 nm. The HWHM of the reflected light is much broader than the earlier transmission cone, having a value of 21.8$^\circ$, with 95.7 $\%$ of normally incident light being reflected. Minimum reflection occurs at an angle of 56.6 degrees, which constitutes the pseudo-Brewster angle for this particular structure. The reflection goes to unity at grazing angles, as one would expect. From the spectral dependence we can again see strong wavelength selectivity, and the transmission window (low reflection area below the red dashed line in the lower part of Fig. 4) is slightly redshifted. Again, for TE polarized light we see a much broader 31.3 degree reflection cone, with an additional side maximum at 46.2$^\circ$ incidence angle, and the pseudo-Brewster angle has shifted to 33.4 degrees.

The reasons behind the angular narrowing and increased transmission/reflection are quite apparent. Normally incident light couples to the Fabry-P\'erot transmission (or reflection) window of the structure efficiently, whereas off axis light does not. This effect is most prominently seen in  Fig.~\ref{reflection_TE}. Light that is incident at an appropriate angle to the surface of the device may couple to guided modes between the layers, and excite plasmons at the dielectric-ENZ interfaces when the input light is TM polarized, explaining the large difference between the two polarizations. Additionally, there are special plasmonic modes present whenever the dielectric constant of the material is low, called the Ferrell-Berreman (FB) modes \cite{FB}, which further absorb light. The FB modes have a relatively large bandwidth, and they can be excited from free space, unlike most plasmon modes.

\begin{figure}[H]
\centering
\includegraphics[width=0.9\columnwidth]{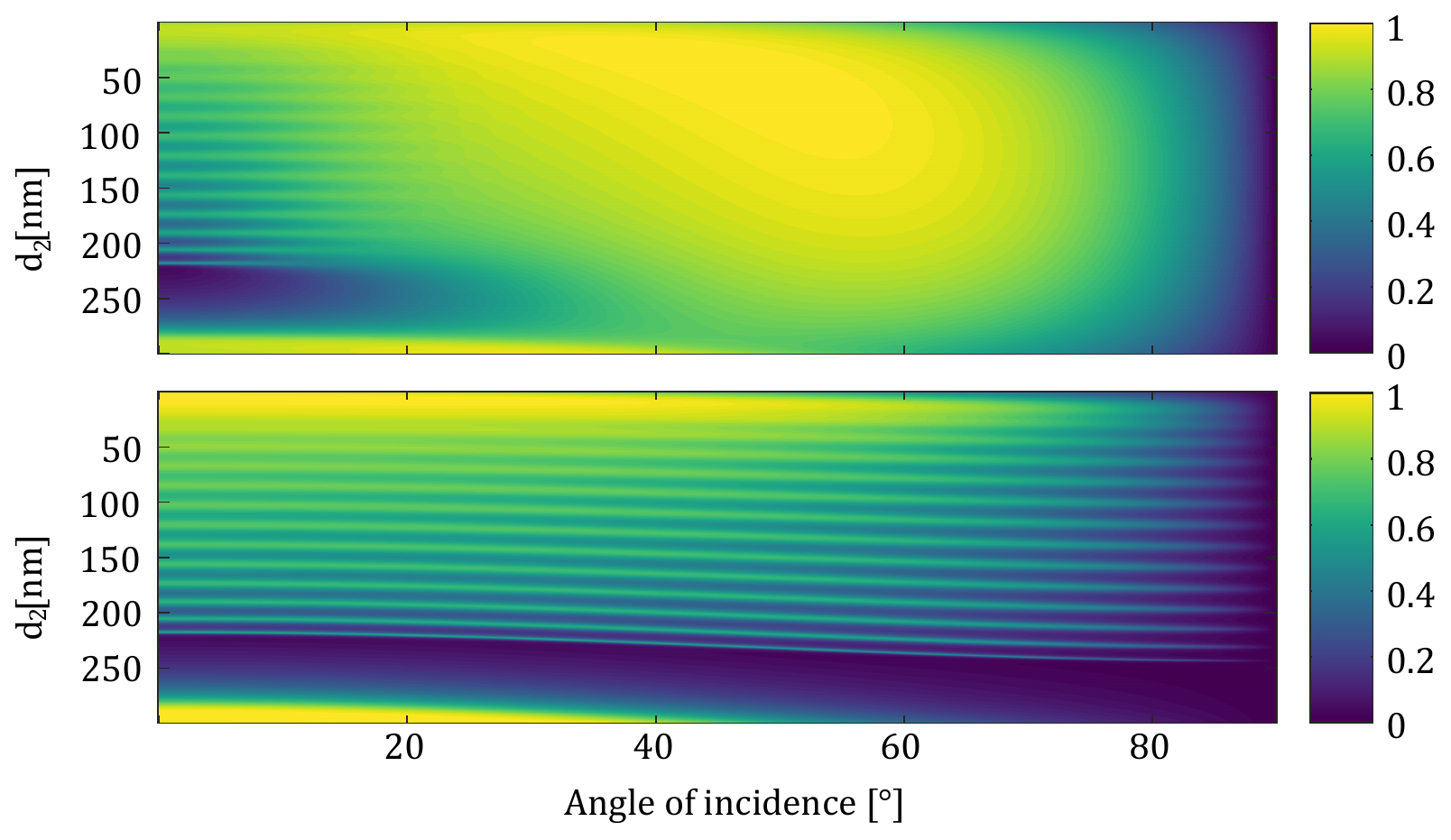}
\caption{Absorption in transmission and reflection optimized structures, top: for TM polarized light as a function of dielectric film thickness, $d_2$, bottom: for TE polarized light as a function of dielectric film thickness, $d_2$. Both have 16 layers of ITO with a thickness of $d_1=64$ nm, and 15 layers of TiO$_2$}
\label{plasmon}
\end{figure}

Since the optimum ITO thickness for both transmission and reflection is 64 nm, the absorption of both of these structures can be depicted in a single ($\theta_i,d_2$)-map, as in Fig.~\ref{plasmon}. The losses caused by plasmon excitations are located within the uniform absorption region above $\sim$30 degree incidence angle, showing a stark difference between TM and TE polarizations. Employing these loss mechanisms leads to the possibility of designing a strongly absorbing multilayer structure, which exhibits a large angular bandwidth.

\subsection{Absorption}

\begin{figure}[H]
\centering
\includegraphics[width=0.9\columnwidth]{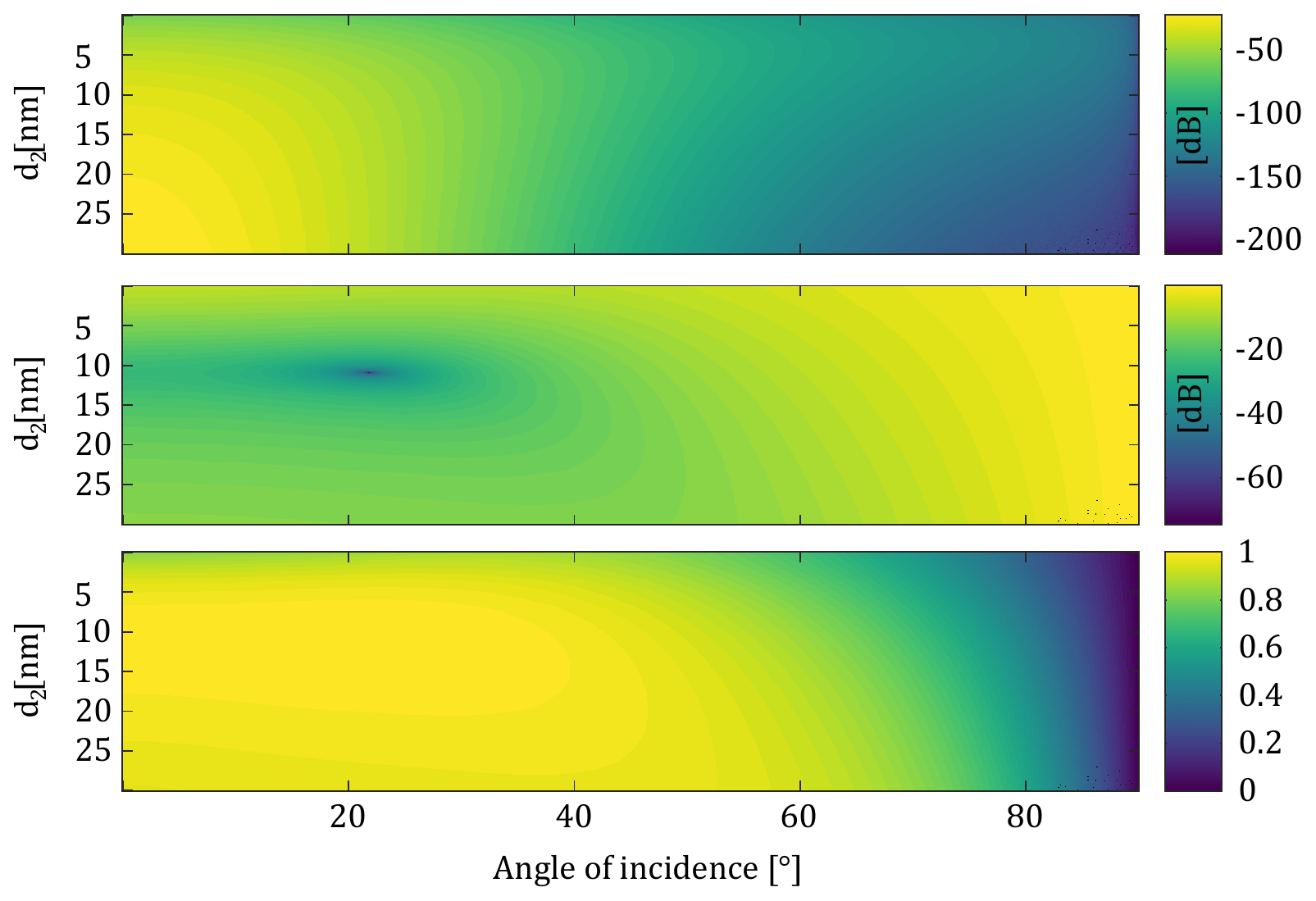}
\caption{Incident light is TM polarized, from top to bottom: transmission, reflection, and absorption, as functions of angle of incidence, $\theta_i$, and dielectric film thickness, $d_2$, with the optimal value being $d_2=11$ nm. There are 59 layers of ITO with a thickness of $d_1=51$ nm, and 60 layers of TiO$_2$. Note that the scale for transmission and reflection is in decibels, whereas the absorption is scaled linearly.}
\label{absorption_map}
\end{figure}

Applying the same parameter sweep as before, but aiming to maximize absorption, we first determined that a thicker stack of 60 layers of TiO$_2$ and 59 layers of ITO -- with dielectric as the first and last films -- produces the highest levels of absorption. Furthermore, we found that the optimum ITO film thickness is 51 nm. The final $(\theta_i,d_2)$ maps for transmission, reflection, and absorption are found in Fig.~\ref{absorption_map}, from top to bottom, respectively. Note that the scale for transmission and reflection is in decibels, whereas the absorption is presented in linear scale.

\begin{figure}[H]
\centering
\includegraphics[width=0.9\columnwidth]{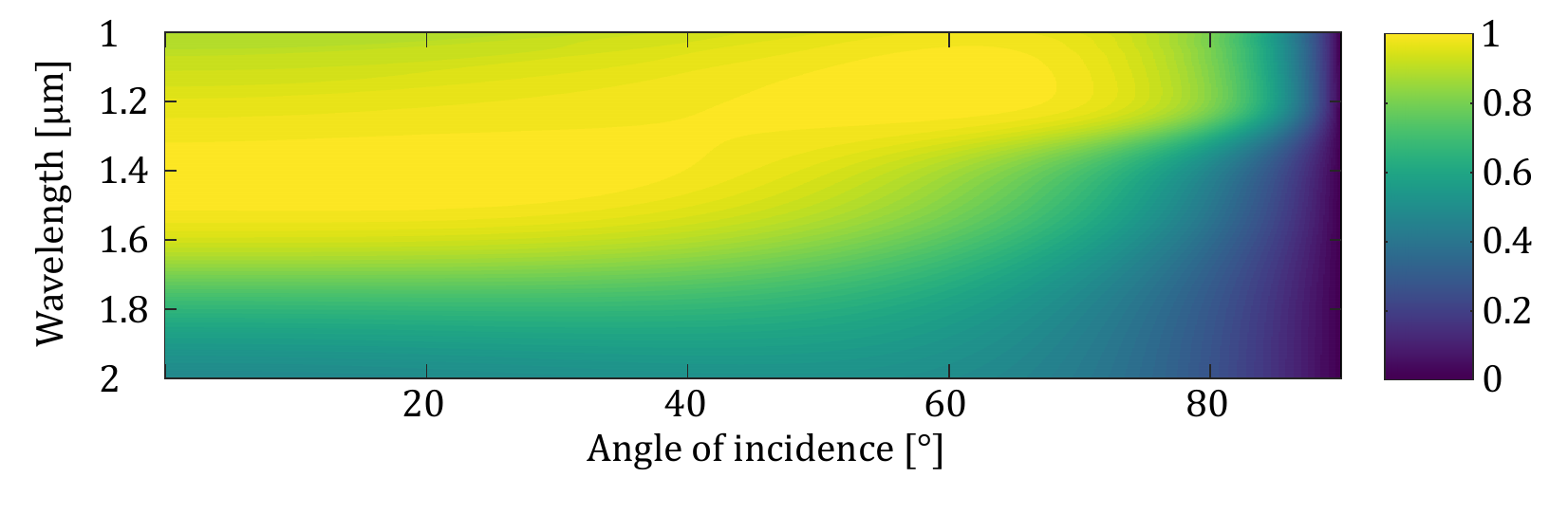}
\caption{Absorption of TM polarized light as a function of wavelength and angle of incidence. There are 59 layers of ITO with a thickness of $d_1=51$ nm, and 60 layers of TiO$_2$ with a thickness of $d_2=11$ nm.}
\label{absorption_wavelength}
\end{figure}

From the optimization cycle, the highest absorption was found at a TiO$_2$ layer thickness of 11 nm. It is quite clear that this value does not present the highest loss in transmitted light, but it is the minimum for reflection. With the selected values for film thicknesses, the reflection is attenuated about -61.5 dB at an angle of 21.8$^\circ$, which constitutes the pseudo-Brewster angle for this structure. A relatively large portion, 99.6 $\%$, of normally incident light is absorbed, and the absorption maximum of 99.998 $\%$ is attained at an angle of 21.9 degrees. Absorption remains above 99.5~$\%$ until the angle of incidence exceeds 30.3$^\circ$, after which reflection starts to rise. The HWHM of the absorption is 77.5$^\circ$, and at larger angles the stack behaves like a grazing angle mirror, leading to extremely poor transmission at steeper incidence angles. The spectral dependence of the structure is depicted separately, in Fig.~\ref{absorption_wavelength}.

As depicted in Fig.~\ref{absorption_wavelength}, these choices lead to strong absorption over a relatively large spectral bandwidth, and the absorption remains above 99 $\%$ from 1353 nm to about 1493 nm (a width of $\Delta \lambda = 140$ nm). The absorption bandwidth is mainly governed by the dispersion relation of the ENZ material, and it is possible to engineer it for specific applications. The plasmon contribution is now blueshifted and at high angles of incidence, the maximum plasmon absorption is centered near 1200 nm.

\begin{figure}[H]
\centering
\includegraphics[width=0.9\columnwidth]{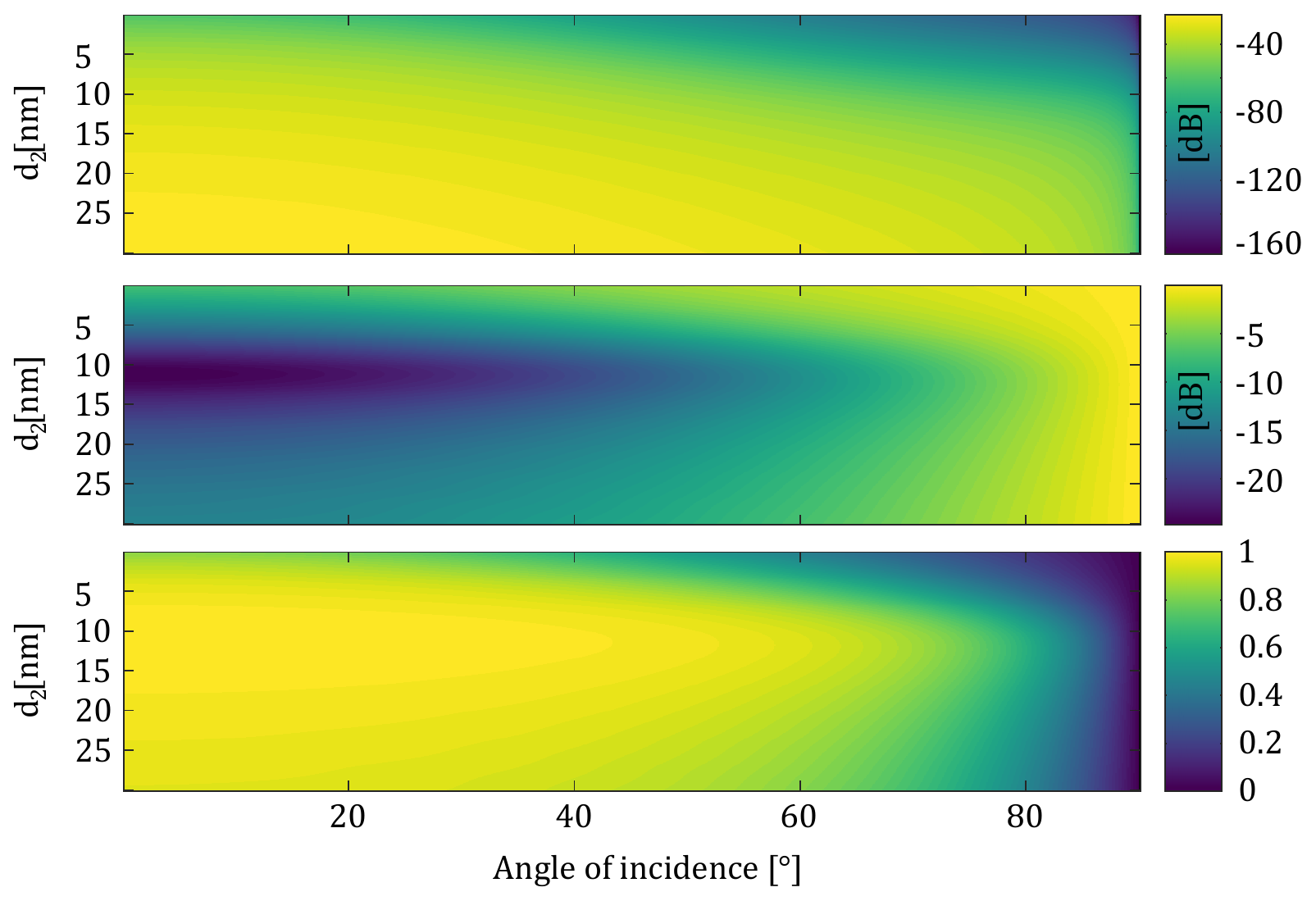}
\caption{Incident light is TE polarized, from top to bottom: transmission, reflection, and absorption, as functions of angle of incidence, $\theta_i$, and dielectric film thickness, $d_2$. There are 59 layers of ITO with a thickness of $d_1=51$ nm, and 60 layers of TiO$_2$. Note that the scale for transmission and reflection is in decibels, whereas the absorption is scaled linearly.}
\label{absorption_map_TE}
\end{figure}

\begin{figure}[H]
\centering
\includegraphics[width=0.9\columnwidth]{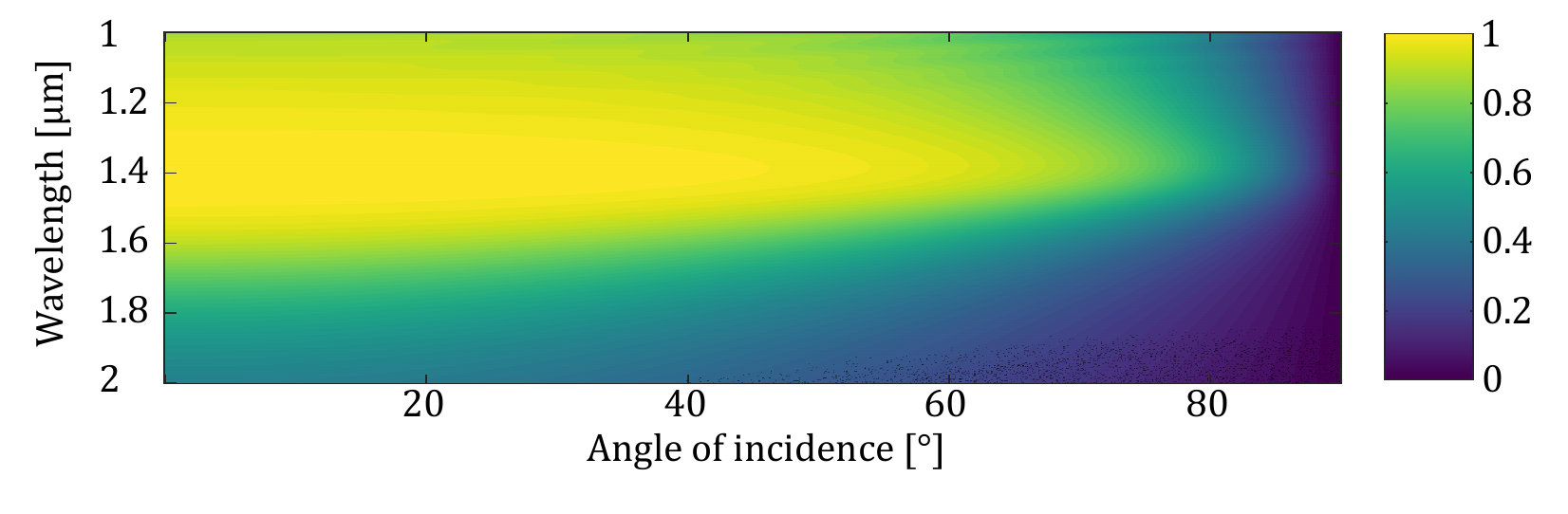}
\caption{Absorption of TE polarized light as a function of wavelength and angle of incidence. There are 59 layers of ITO with a thickness of $d_1=51$ nm, and 60 layers of TiO$_2$}
\label{absorption_wavelength_TE}
\end{figure}

The corresponding transmission, reflection and absorption maps for TE polarized input light are found in Fig.~\ref{absorption_map_TE}, and the spectral dependence is in Fig.~\ref{absorption_wavelength_TE}. Now the telltale signs of TM polarized light are gone, as there are no plasmon or Brewster's angle contributions. Therefore, the highest absorption is achieved at normal incidence, with a value of 99.6 $\%$, and a HWHM of 82.8 degrees.

\section{Discussion and conclusions}

The above examples are rather extreme, but they use existing materials and can be fabricated with current methods -- at least in principle. The properties of ENZ materials can be leveraged in thin film stack design, which offers a lot of freedom in designing structures for multiple purposes. Although possible, fabricating these stacks may prove to be an experimental challenge, since the resonances attained at different dielectric film thicknesses are only a few nanometers apart. More importantly, the resonance peaks are very sharp with respect to the dielectric thickness, and a working experimental realization will need to employ a resonance peak that is not as sharp, if the layer thickness cannot be exactly controlled. This will no doubt degrade the performance of the structure, but it is still a large improvement over bulk ENZ materials that are currently available. The wavelength tunability might alleviate the demands on fabrication, given that there is a precisely controlled light source available at the desired wavelength range. 

The modeling method does not assume that the stack is an effective medium, and the multiple reflections and even plasmon excitations are taken into account. The modeled examples are experimentally realizable, and constitute large improvements to the reflective, refractive, or absorptive properties of ENZ materials. For example, the angular spread of transmission can be reduced to only a few degrees while the transmission efficiency is increased nearly 26 fold. On the other hand, absorption can be maximized to levels seen only in novel surface patterned absorbers. The only limiting factor in the discussed structures is the performance of the ENZ media, and a smaller imaginary part of the dielectric function would improve the operation of the transmitting and reflecting structures. For example, if the ENZ had a permittivity of $\epsilon_{1} = i\,0.1$, then the transmission HWHM would be 1.48$^\circ$, and 71.6 $\%$ of normally incident light would be transmitted. This corresponds to a low index material with $n=0.0007$, although the ENZ index would be $n_2 = 0.22$. Similarly, for the reflecting stack the HWHM would be 9.3$^\circ$, and 99.1$\%$ of normally incident light would be reflected. These structures can be used to realize angular pinholes that either transmit or reflect only in a certain cone, which can be fairly small. A narrow enough angular pinhole would be useful in many applications, such as angular signal discrimination, directional light generation and development of high quality laser light sources. Notably, it would be possible to create lasers with almost arbitrary cavity shapes, as long as the cavity mirrors allow oscillation only in a tight beam. 

The anomalously strong absorption is remarkable in the sense that there is no surface modification, and all of the materials are perfectly planar layers. Even though this is the case, the metamaterial we have designed here rivals the absorbing properties of the darkest materials ever made, such as VANTABLACK, or carbon nanotube--metal hierarchical architectures, which feature minimum reflectivities of -38 dB and -50 dB, respectively \cite{blackest}. Although the minimum reflectance of -61.5 dB of our structure is far smaller than either of these values, it can be achieved only at a specific angle of incidence, and the angular and spectral bandwidths are not as wide. The angular bandwidth could be increased with surface patterning, but to engineer the spectral bandwidth one would require a new ENZ. Materials that feature either flatter or steeper dispersion curves could be used to tune the spectral absorption bandwidth. The absorbing structures show promise for novel attenuators for use in thermal photovoltaics, photodetection, or emissivity control, for example.

It is be possible to realize spatial coherence control and coherence switching with these structures, which we will report on separately. Another fairly interesting direction for our future research is the possibility of adding optical gain into the system, for example in the form of quantum dots embedded into the stack.

\section*{Acknowledgements}

We acknowledge the financial support of Academy of Finland Flagship Programme (PREIN), (320165, 320166), Academy of Finland projects (322002, 310511), and Academy of Finland  Competitive funding to strengthen university research profiles (301820).

\section*{References}

\bibliography{references}

\providecommand{\newblock}{}
\begin{thebibliography}{10}
\expandafter\ifx\csname url\endcsname\relax
  \def\url#1{{\tt #1}}\fi
\expandafter\ifx\csname urlprefix\endcsname\relax\def\urlprefix{URL }\fi
\providecommand{\eprint}[2][]{\url{#2}}

\bibitem{7Hajian}
Hajian H, Ozbay E and Caglayan H 2017 {\em Scientific Reports\/} {\bf 7} 4741
  ISSN 2045-2322 \urlprefix\url{https://doi.org/10.1038/s41598-017-04680-y}

\bibitem{8Hajian}
Hajian H, Ozbay E and Caglayan H 2016 {\em Applied Physics Letters\/} {\bf 109}
  031105 (\textit{Preprint} \eprint{https://doi.org/10.1063/1.4959085})
  \urlprefix\url{https://doi.org/10.1063/1.4959085}

\bibitem{9Feng}
Feng S and Halterman K 2012 {\em Phys. Rev. B\/} {\bf 86}(16) 165103
  \urlprefix\url{https://link.aps.org/doi/10.1103/PhysRevB.86.165103}

\bibitem{10Alu2007}
Al\`u A, Silveirinha M~G, Salandrino A and Engheta N 2007 {\em Phys. Rev. B\/}
  {\bf 75}(15) 155410
  \urlprefix\url{https://link.aps.org/doi/10.1103/PhysRevB.75.155410}

\bibitem{11Wurtz}
Wurtz G~A, Pollard R, Hendren W, Wiederrecht G~P, Gosztola D~J, Podolskiy V~A
  and Zayats A~V 2011 {\em Nature Nanotechnology\/} {\bf 6} 107--111 ISSN
  1748-3395 \urlprefix\url{https://doi.org/10.1038/nnano.2010.278}

\bibitem{Alam}
Alam M~Z, De~Leon I and Boyd R~W 2016 {\em Science\/} {\bf 352} 795--797 ISSN
  0036-8075 (\textit{Preprint}
  \eprint{https://science.sciencemag.org/content/352/6287/795.full.pdf})
  \urlprefix\url{https://science.sciencemag.org/content/352/6287/795}

\bibitem{Alireza}
Rashed A~R, Yildiz B~C, Ayyagari S~R and Caglayan H 2020 {\em Phys. Rev. B\/}
  {\bf 101}(16) 165301
  \urlprefix\url{https://link.aps.org/doi/10.1103/PhysRevB.101.165301}

\bibitem{realandimag}
Javani M~H and Stockman M~I 2016 {\em Phys. Rev. Lett.\/} {\bf 117}(10) 107404
  \urlprefix\url{https://link.aps.org/doi/10.1103/PhysRevLett.117.107404}

\bibitem{zim}
Liberal I and Engheta N 2017 {\em Nature Photonics\/} {\bf 11} 149--158 ISSN
  1749-4893 \urlprefix\url{https://doi.org/10.1038/nphoton.2017.13}

\bibitem{stratified}
Yeh P 2005 {\em Optical waves in layered media\/} (Wiley)

\bibitem{Humeyra}
Caglayan H, Hong S~H, Edwards B, Kagan C~R and Engheta N 2013 {\em Phys. Rev.
  Lett.\/} {\bf 111}(7) 073904
  \urlprefix\url{https://link.aps.org/doi/10.1103/PhysRevLett.111.073904}

\bibitem{directional}
Enoch S, Tayeb G, Sabouroux P, Gu\'erin N and Vincent P 2002 {\em Phys. Rev.
  Lett.\/} {\bf 89}(21) 213902
  \urlprefix\url{https://link.aps.org/doi/10.1103/PhysRevLett.89.213902}

\bibitem{Moitra2013}
Moitra P, Yang Y, Anderson Z, Kravchenko I~I, Briggs D~P and Valentine J 2013
  {\em Nature Photonics\/} {\bf 7} 791--795 ISSN 1749-4893
  \urlprefix\url{https://doi.org/10.1038/nphoton.2013.214}

\bibitem{review}
Niu X, Hu X, Chu S and Gong Q 2018 {\em Advanced Optical Materials\/} {\bf 6}
  1701292 (\textit{Preprint}
  \eprint{https://onlinelibrary.wiley.com/doi/pdf/10.1002/adom.201701292})
  \urlprefix\url{https://onlinelibrary.wiley.com/doi/abs/10.1002/adom.201701292}

\bibitem{FB}
Vassant S, Hugonin J~P, Marquier F and Greffet J~J 2012 {\em Opt. Express\/}
  {\bf 20} 23971--23977
  \urlprefix\url{http://www.opticsexpress.org/abstract.cfm?URI=oe-20-21-23971}

\bibitem{blackest}
Cui K and Wardle B~L 2019 {\em ACS Applied Materials \& Interfaces\/} {\bf 11}
  35212--35220 pMID: 31514497 (\textit{Preprint}
  \eprint{https://doi.org/10.1021/acsami.9b08290})
  \urlprefix\url{https://doi.org/10.1021/acsami.9b08290}

\end{thebibliography}

\end{document}